High Energy pp Elastic Scattering in Condensate Enclosed Chiral Bag Model
And
TOTEM Elastic Measurements at LHC at 7 TeV

M. M. Islam[a], R. J. Luddy[b]

Department of Physics, University of Connecticut, Storrs, CT 06269 USA
a) islam@phys.uconn.edu b) rjluddy@phys.uconn.edu

## Abstract

We study high energy pp and p̄p elastic scattering in the TeV region based on an effective field theory model of the proton. We phenomenologically investigate the main processes underlying elastic scattering and quantitatively describe the measured elastic $d\sigma/dt$ at energies 7.0 TeV (LHC pp), 1.96 TeV (Tevatron p̄p), and 0.630 TeV (SPS p̄p). Finally, we give our prediction for pp elastic $d\sigma/dt$ at 14 TeV – that will be measured by the TOTEM Collaboration.

High energy pp and p̄p elastic scattering came to the forefront of particle physics research with the advent of the pp and p̄p colliders: 1) CERN pp ISR Collider in the early seventies at energy ~ 50 GeV; 2) CERN p̄p SPS Collider in the mid-eighties at energy ~0.5 TeV; 3) Fermilab p̄p Tevatron Collider in the early nineties at energy 1.8 TeV; and 4) our present CERN LHC where elastic pp scattering has been recently measured at energy 7 TeV by the TOTEM Collaboration.

My collaborators and I have been studying high energy pp and p̄p elastic scattering over a period of more than three decades beginning in the early eighties. Our study has two distinct aspects. One is to build an effective field theory model of the proton[1, 2)]. The other is to carry out phenomenological investigation of pp and p̄p elastic scattering in the TeV region with the goals: i) identify the main processes underlying elastic scattering based on our field theory model of the proton; ii) describe quantitatively the measured elastic $d\sigma/dt$ at 7.0 TeV (pp), 1.96 TeV (p̄p), and 0.630 TeV (p̄p); iii) predict pp elastic $d\sigma/dt$ at LHC at 14 TeV—which is planned to be measured by the TOTEM Collaboration.

The combined theoretical development and phenomenological investigation have led us to a physical picture of the proton (Fig. 1)[3)]. The proton has three regions: an outer region consisting of a quark-antiquark (qq̄) condensed ground state; an inner shell of baryonic charge — where the baryonic charge is topological (or, geometrical) in nature (similar to the "Skyrmion Model" of the nucleon); and a core region of size 0.2 fm, where the valence quarks are confined. The part of the proton structure comprised of a shell of baryonic charge and three valence quarks in a small core has been known as a "Chiral Bag" model of the nucleon in low-energy studies[4)]. What we are finding from high energy elastic scattering then is that the proton is a "Condensate Enclosed Chiral Bag".

The proton structure shown in Fig. 1 leads to three main processes in elastic scattering as shown in Fig. 2. The first process occurs in the small $|t|$ region, i.e. in the near forward direction, where the outer cloud of qq̄ condensate of one proton interacts with that of the other giving rise to diffraction scattering. The latter underlies the observed increase of the pp total cross section with energy as $(\ln s)^2$ and equal pp and $\bar{\mathrm{p}}$p total cross sections at high energy.

The second process comes into play at $|t| \gtrsim 0.5$ GeV$^2$, when the topological baryonic charge of one proton probes that of the other via vector meson $\omega$-exchange. This process is analogous to one electric charge probing another via photon exchange. The spin-1 $\omega$ acts like a photon, because of its coupling with the topological baryonic charge.

The third process also occurs at $|t| \gtrsim 0.5$ GeV$^2$ (transverse distance $b \lesssim 0.3$ fm, $b \sim \frac{1}{q}$), when elastic scattering originates from the hard collision of a valence quark of one proton with that of the other. This process can be better visualized in momentum space (Fig. 3).

Our recent investigations have shown that, because of the high c.m. energy of the LHC (7 TeV), multiple $\omega$-exchanges have to be taken into account in pp (and $\bar{\mathrm{p}}$p) elastic scattering. Furthermore, each $\omega$-exchange is accompanied by a glancing collision between a layer of scalar particles of one proton with that of the other (Fig. 4). The layer of scalar particles that envelops the baryonic charge shell (Fig.1) of a proton originates from a scalar field present in our proton field theory model. As far as valence qq scattering is concerned (Fig.3), we find that it is due to (crossing-odd) low-x gluon-gluon interaction. Furthermore, it, too, is accompanied by the glancing collision between one layer of scalar particles of a proton with that of the other

With the above phenomenological features taken into account, we are able to describe quantitatively:
i) pp elastic scattering at LHC at 7.0 TeV measured by the TOTEM Collaboration[5,6)];
ii) $\bar{\mathrm{p}}$p elastic scattering at the Tevatron at 1.96 TeV measured by the D0 Collaboration[7)];
iii) the earlier $\bar{\mathrm{p}}$p elastic scattering at the SPS Collider at 0.630 TeV measured by the UA4 Collaboration[8)]. The $d\sigma/dt$ calculated by us for these energies are shown in Fig. 5.

Finally, we show in Fig. 5 our prediction of pp elastic scattering at LHC at 14 TeV. If the planned measurement of elastic $d\sigma/dt$ at LHC at 14 TeV by the TOTEM Collaboration shows satisfactory quantitative agreement with our prediction for 14 TeV, then the underlying picture of the proton shown in Fig. 1 as a Condensate Enclosed Chiral Bag will be confirmed.

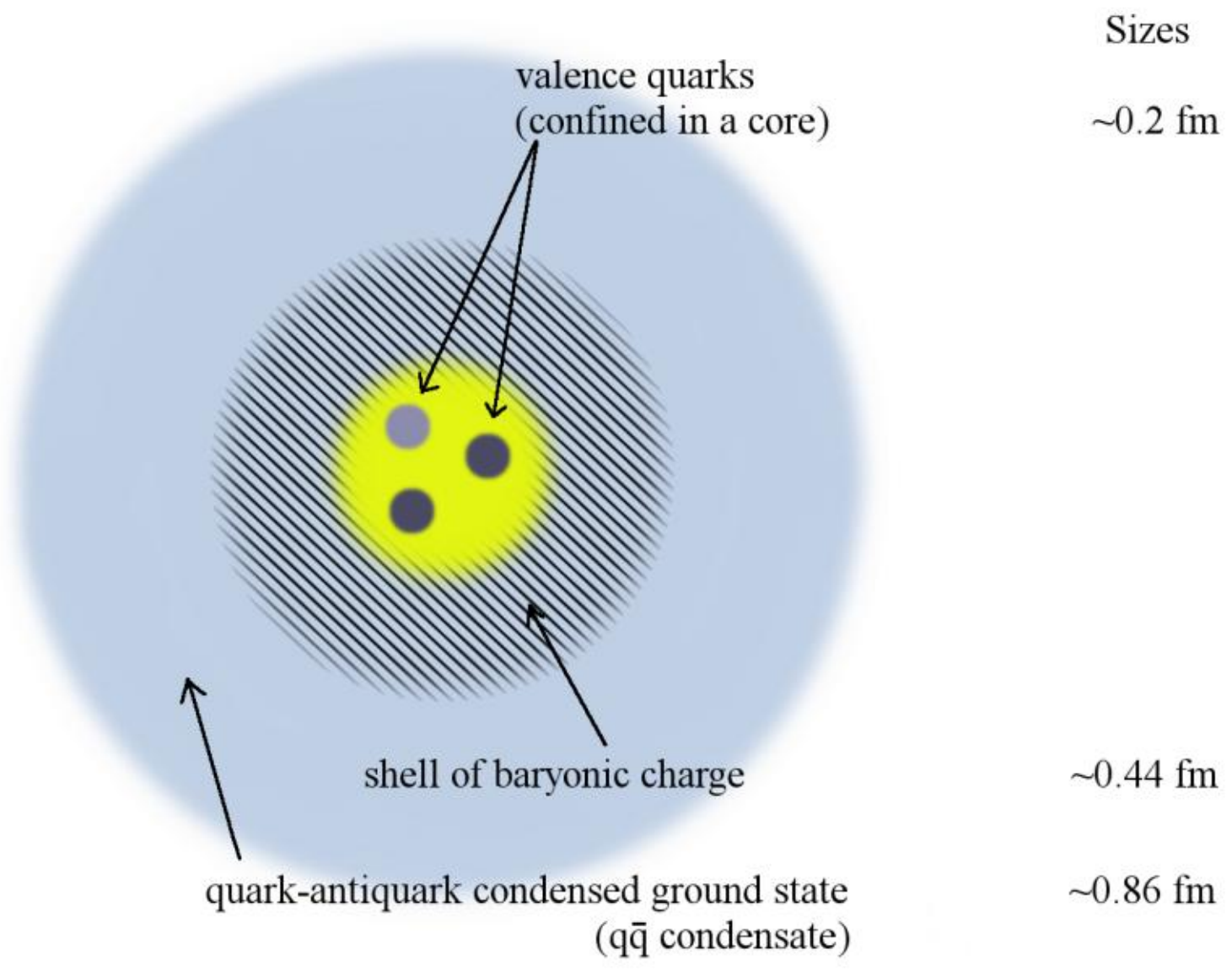

Fig. 1. Physical picture of the proton as a Condensate Enclosed Chiral Bag.

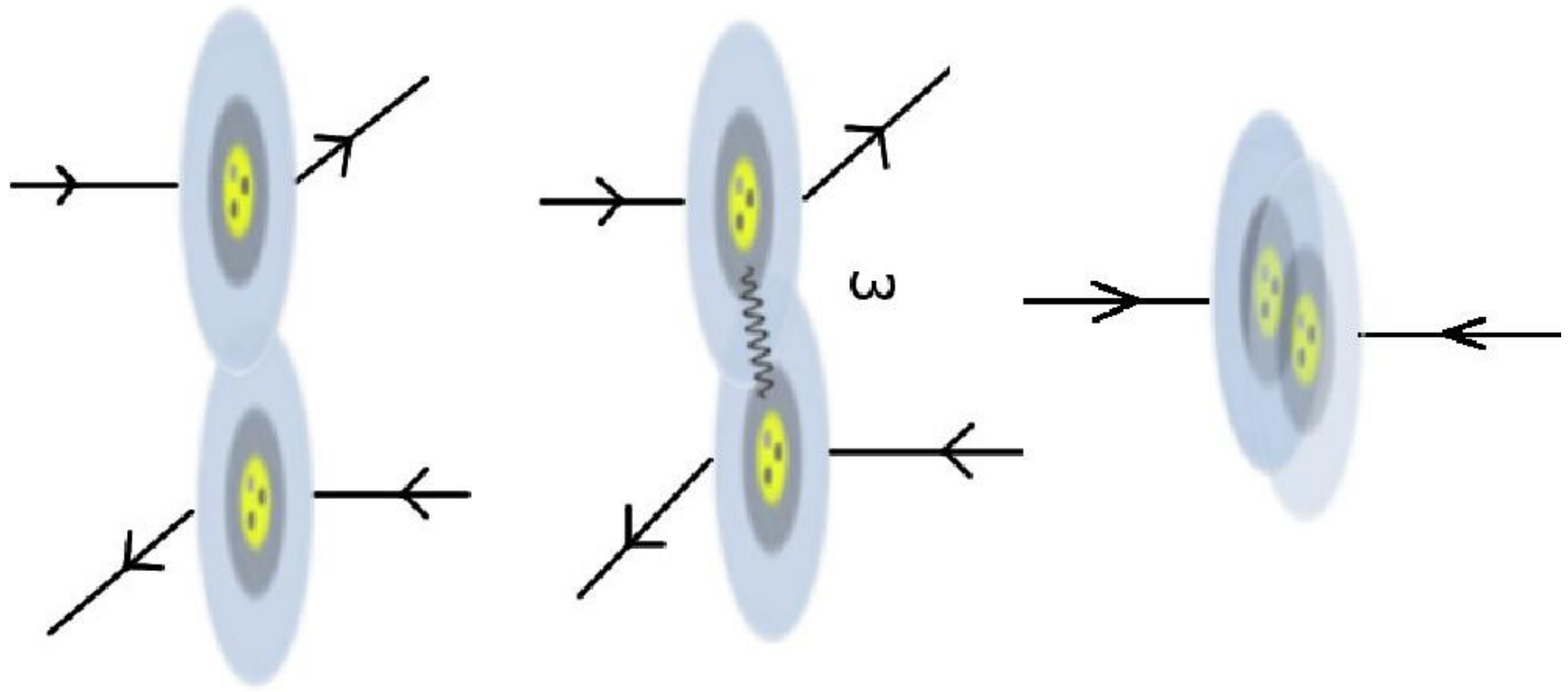

Fig. 2. Elastic scattering processes (from left to right):
1) diffraction, 2) $\omega$-exchange, 3) short-distance collision ($b \lesssim 0.3$ fm).

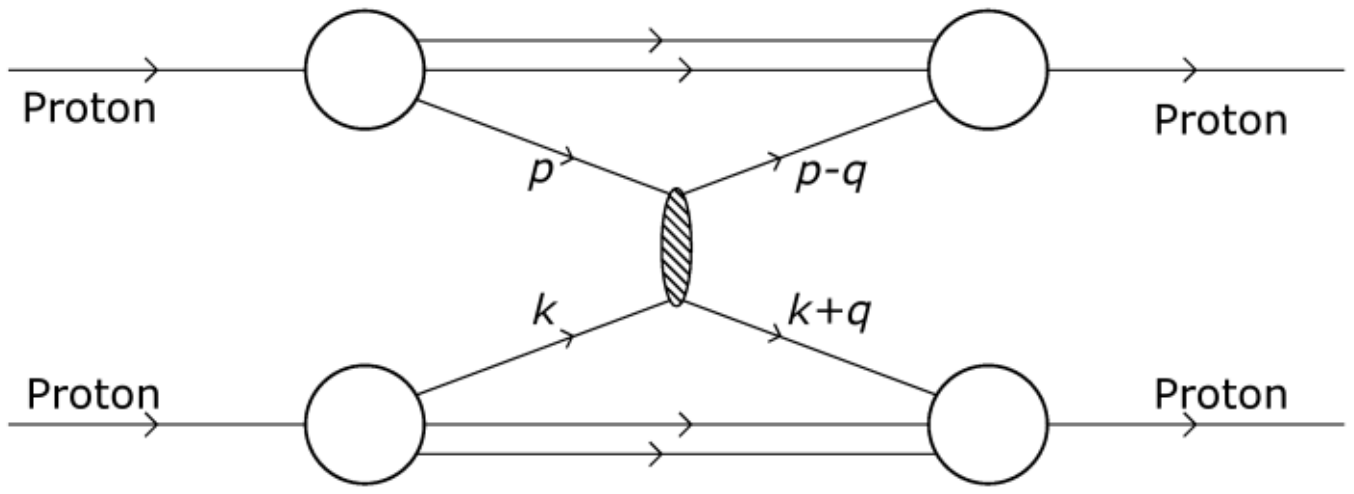

Fig. 3. Hard collision of a valence quark of 4-momentum $p$ from one proton with a valence quark of 4-momentum $k$ from the other proton, where the collision carries off the whole momentum transfer $q$.

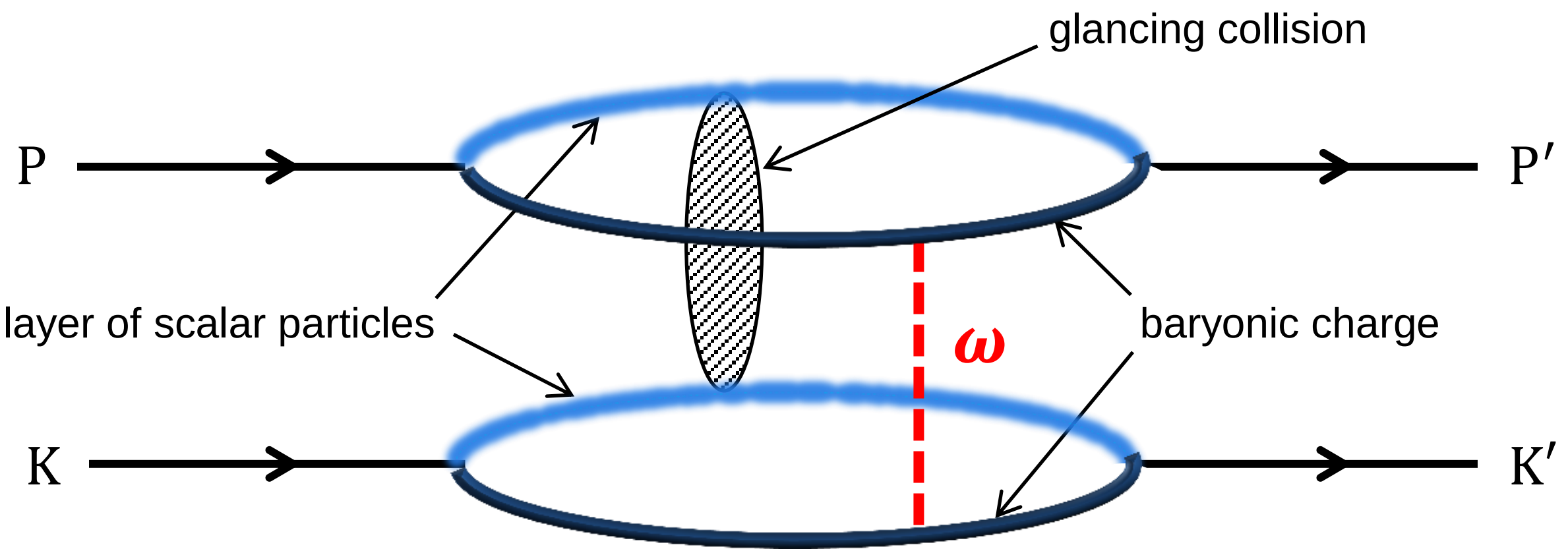

Fig. 4. Single $\omega$-exchange accompanied by glancing collision of a layer of scalar particles of one proton with that of the other (in momentum space).

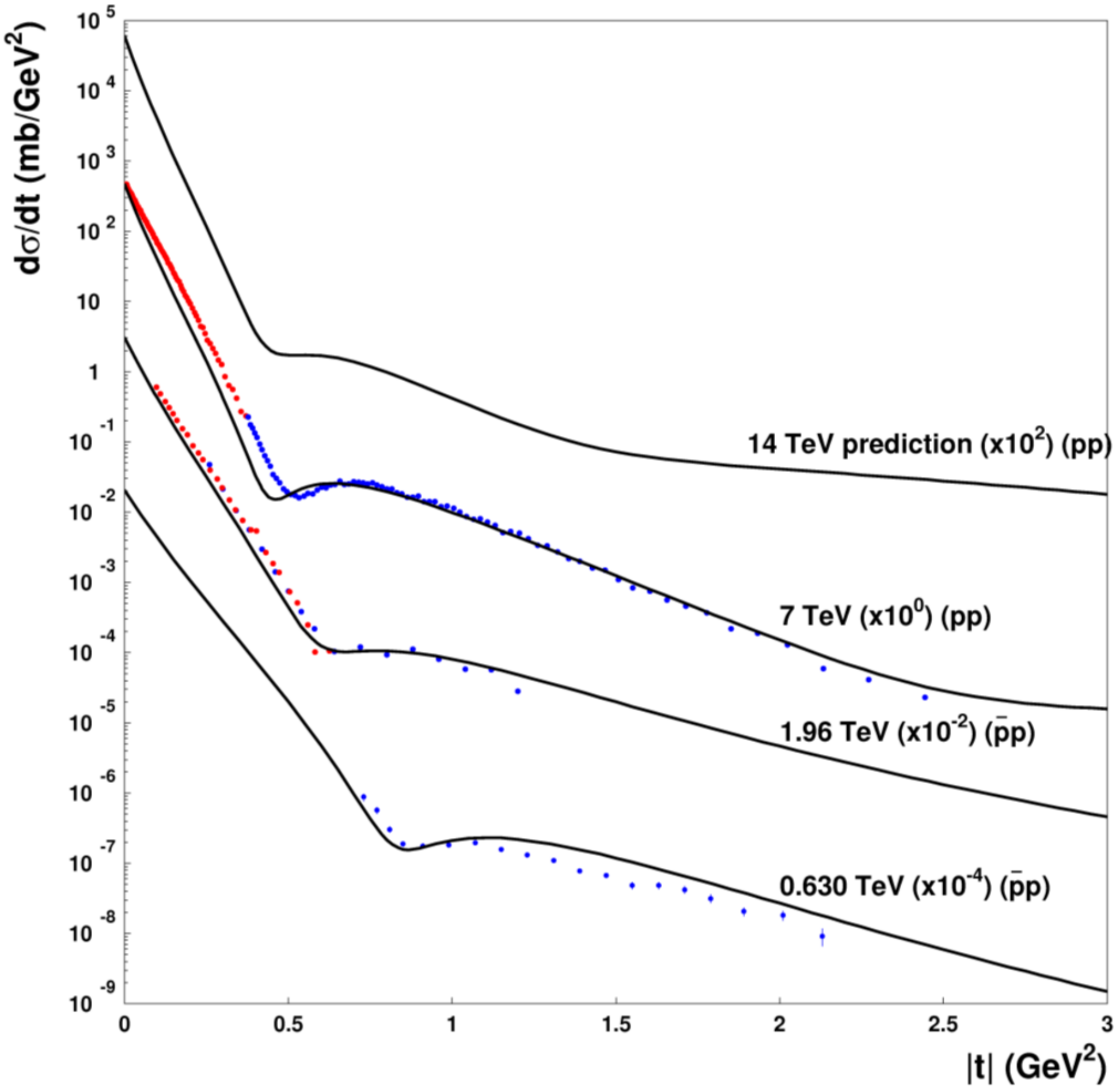

Fig. 5. Our calculated differential cross sections at 7.0 TeV (pp), 1.96 TeV (p̄p), and 0.630 TeV (p̄p) are shown (black curves). The experimental data are a) pp measurements at 7.0 TeV by TOTEM Collaboration (red and blue dots)[5, 6]; b) p̄p measurements at 1.96 TeV by D0 Collaboration (blue dots)[7] together with CDF and E710 measurements at 1.80 TeV (red dots); c) p̄p measurements at 0.630 TeV by UA4 Collaboration[8].
The black curve for 14 TeV $d\sigma/dt$ is our prediction for pp elastic scattering, which will be measured at LHC by the TOTEM Collaboration.